# Robust Image Watermarking Under Pixel Wise Masking Framework

V. H. MANKAR, T. S. DAS, S. SAHA AND S. K. SARKAR
Dept. of Electronics & Telecommunications, Jadavpur University, Kolkata, India
vijaymankar@yahoo.com, tirthasankardas@yahoo.com, sjit.saha@gmail.com,
su_sircir@yahoo.co.in,

**Abstract**
*The current paper presents a robust watermarking method for still images, which uses the similarity of discrete wavelet transform and human visual system (HVS). The proposed scheme makes the use of pixel wise masking in order to make binary watermark imperceptible to the HVS. The watermark is embedded in the perceptually significant, spatially selected detail coefficients using sub band adaptive threshold scheme. The threshold is computed based on the statistical analysis of the wavelet coefficients. The watermark is embedded several times to achieve better robustness. Here, a new type of non-oblivious detection method is proposed. The improvement in robustness performance against different types of deliberate and non-intentional image impairments (lossy compression, scaling, cropping, filtering etc) is supported through experimental results. The reported result also shows improvement in visual and statistical invisibility of the hidden data. The proposed method is compared with a state of the art frequency based watermarking technique, highlighting its performance. This algorithmic architecture utilizes the existing allocated bandwidth in the data transmission channel in a more efficient manner.*

## 1. Introduction

The last decade has witnessed the rapid development in information technologies that has improved the ease of access to digital information. This leads to the problem of illegal copying and redistribution of digital media. The concept of digital watermarking came in order to solve the problems related to the intellectual property of media. The watermarking embeds a signal into the host data in some invisible way that is supposed to identify the owner [1-4]. Important properties of an image watermarking system are imperceptibility (the watermarking process should not degrade the image significantly), robustness (resistance of the mark against intentional or unintentional attacks like AWGN, filtering, lossy compression, scaling, cropping), data hiding capacity (the amount of information that can be embedded into the original cover work without causing serious distortions) and computational cost and complexity. Moreover, there are several other criteria that can be used to classify watermarking systems such as

- The selection of the locations where the mark is embedded using human visual models, or a randomly generated key,

- The watermarking domain: spatial domain or transform domain. Some of the transforms are the Discrete Cosine Transform (DCT), the Discrete Wavelet Transform (DWT), and the Discrete Fourier Transform (DFT), Fourier-Mellin Transform, the Complex Wavelet Transform (CWT), or Ridgelet Transform [7],

- Encoding of payload: using LSB Modulation, HVS, spread spectrum (SS) & Quantization Index Modulation (QIM) techniques and/or error correction codes (ECC),

- Formation of the watermarked signal: additive, multiplicative or quantization-based,

-The watermark decoder: oblivious (the cover work is not known at the decoder, only the secret key is used), semi-blind (using the watermarked data and the secret key) or non-oblivious (using the cover work and the secret key).

However, these watermarking properties and criteria are related in conflicting manner and the particular algorithmic development emphasizes to a greater extent on one or more such requirements depending on the type of application.

Several papers that deal with copyright protection for images argue that the mark should be embedded in some transform domain, selecting only perceptually significant coefficients, because those are the most likely to survive compression. Cox *et al.* embeds a continuous watermark in the largest 1000 DCT coefficients of the original image, except the DC coefficient, thus spreading its energy on several bins of frequency [2]. Detection is made using the similarity between the two watermarks. Xia *et al.* [5] insert several watermarks in the DWT domain in each detail image, except the approximation sub band, suggesting that the detection could be done hierarchically, computing cross-correlations of the watermark and the difference between the two images for each resolution level. We propose a technique that embeds the watermark into perceptually significant wavelet coefficients using pixel wise masking. The watermark is embedded repeatedly into the detail sub bands, thus increasing the robustness of the method.

The paper is organized as follows: The section 2 presents the watermark embedding and detection in the present work. Section 3 shows the experimental results on 16 X 16 binary watermark and finally section 4 concludes and remarks about some of the aspects analyzed in this paper.

## 2. Watermark Embedding and Detection

Here we explain the chosen method for embedding the watermark. The watermark is embedded into the detail sub bands of the cover image of size 256 X 256. A binary sequence $B_i$, $i \in \{-1, 1\}$, of figure 'M' of size 16X16 is being used as a watermark. The original image is decomposed by $L$ levels using Discrete Wavelet Transform (DWT) through Daubechies 2pt analysis wavelet filter bank [6]. The HVS is sensitive for the changes in smooth parts of the image, not for small changes in high frequencies. So the watermark is embedded in each of the 9 sub bands in 3 levels excluding the LL sub band. The algorithm is explained as follows:

The maximum of the coefficients from each sub band are computed as given by the equation:

$$M_{s,l} = \max(\max(C_{s,l})) \quad (1)$$

where $M_{s,l}$ is the maximum of the coefficients from the respective detail image, $C$ is the original cover image, $s \in \{h, v, d\}$ ($h, v, d$ stands for "horizontal", "vertical" and "diagonal", respectively) and $l=1,\ldots,L$. Now for each sub band (except for the LL sub band), the threshold $T_{s,l}$ is computed as:

$$T_{s,l} = q_l . M_{s,l} \quad (2)$$

where $q_l$ is a level-dependent variable.

Here the level-dependent variables are $q_1$ (1st level), $q_2$ (2nd level) and $q_3$ (3rd level). If the wavelet coefficient seems to be less or equal than the computed threshold, the coefficients are left unmodified. The watermark embedding modulation function is mathematically expressed as:

$$c^w_{s,l}(m,n) = c_{s,l}(m,n)[1 + \alpha . b(m,n)] \quad (3)$$

where $c^w_{s,l}(m, n)$ denotes the watermarked image, $c_{s,l}(m, n)$ denotes the DWT coefficients of the original image and α is a parameter, $\alpha \in (0,1)$, controlling the level of the watermark. But using the negative modulation function, watermarking technique is slightly changed as:

$$c^w_{s,l}(m,n) = c_{s,l}(m,n)[1 - \alpha . b(m,n)] \quad (4)$$

The negative sign implies that when the watermarked bit is negative it is added with image coefficient and vice versa. This ensures the better robustness efficiency than being used with the positive modulation.. The higher the strength of the mark α and the lower the parameter $q_l$ are, the more robust yet visible the watermark will be. The watermarked image $C^w$ is thus computed from the newly modified coefficients using the Daubechies 2pt synthesis wavelet filter bank.

The detection process is the inverse procedure of the insertion process. The detection scheme is a non-blind watermarking scheme. So the original image is required in the decoder side to extract the watermark. To detect the watermark from the watermarked distorted image C', first the original and the received images are transformed using DWT using the same analysis filter. Then once again the threshold is calculated from the original image by the same process as in the embedding process.

$$M_{s,l} = \max(\max(C_{s,l})) \quad (5)$$

$$T_{s,l} = q_l . M_{s,l} \quad (6)$$

If $c_{s,l}(m, n) > T_{s,l}$, then we extract the watermark using the equation as:

$$\tilde{b}(m,n) = \text{sgn}\left[\frac{c'_{s,l}(m,n) - c_{s,l}(m,n)}{c_{s,l}(m,n)}\right] \quad (7)$$

Where $\tilde{b}(m,n)$ is the recovered watermark bit. By this way the watermarks are extracted from all 9 sub bands. The final watermark is computed from these 9 sub bands watermarks by making a comparison using the majority rule defined as follows: the most common bit values from the recovered sub bands are assigned to the final watermark. This detector structure is modified by taking various combinations of sub bands in order to improve the robustness performance.

## 3. Results and Discussion

The proposed algorithm is tested over a various number of benchmark images e.g. Lena, Fishing boat, Pepper, Baboon, US Air Force etc. having dimension 256 X 256. The subjective binary watermark of size 16 X 16 with $N_w = 256$ is used for embedding. The Daubechies 2pt (db2) wavelet is used to produce the wavelet coefficients. The various parameter used in the present algorithm are as follows: the number of resolution levels $L = 3$; the strength of the watermark α = 0.4; the level-dependent variables: $q_1 = 0.06$, $q_2 = 0.04$ and $q_3 = 0.02$. The watermark is extracted by a majority rule being used with different types of detector structures. The results obtained by several structures motivated us to select the two best-optimized watermark decoders:

- ❑ Detector I, taking all 9 sub bands in 3 decomposition levels into account.
- ❑ Detector II, taking only a few selective sub bands (horizontal sub band at L=2 and vertical sub bands at L=2 and 3).

The algorithm affects altogether 2304 coefficients from a total of 65536 (including the LL sub band) in such a way that it is still well below the just noticeable distortion (JND) of HVS. The distortion caused by the watermark can be measured by the peak signal-to-noise

69

ratio (PSNR), SSIM respectively [9]. The security value for data hiding can be measured using Kulback Leibler distance [10]. The various transparency and security metrics are supported by given table 1. The experimental results of robustness performances are given in table 2 as well as in fig 1 and 2.

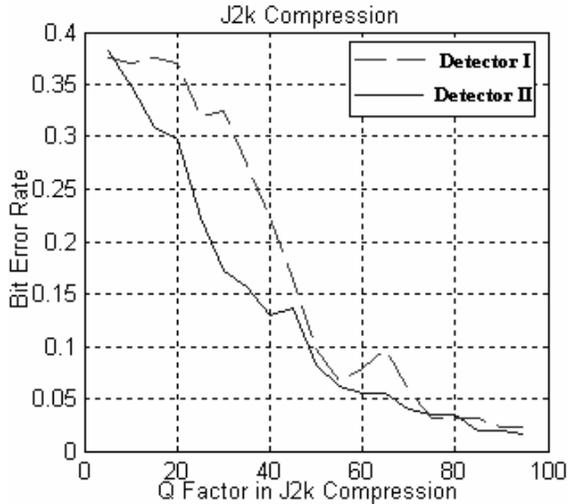

**Fig. 1: Measure of Robustness against JPEG2000 Compression.**

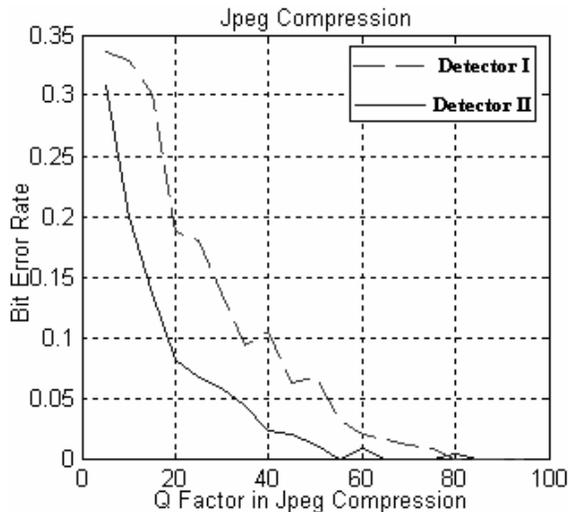

**Fig. 2: Measure of Robustness against JPEG Compression.**

We compare our results with the method proposed by Corina Nafornita keeping the same PSNR values as both use the same sub band adaptive threshold scheme [8]. This comparative performance analysis is supported by the table 3. The proposed method shows better robustness against different image impairments. This is due to the fact that there are architectural differences between our and their algorithm.

We use Daubechies 2pt wavelet and a -ve modulation function having gain α = 0.4 as opposed to their Daubechies 10pt wavelet, +ve modulation function with α = 0.3. A specific subjective watermark is being used as compare to their random data sets. The computational cost and complexity is comparatively high with the Daubechies 10pt wavelet. Moreover, the process of recognition of watermark was in objective manner with an average of 32 results still showing less robustness. The proposed method extracts the watermark image in both subjective and objective manners in single pass with better robustness which is shown in fig. 3.

**Table 1: Results for PSNR, SSIM, Security Values and Mutual Information.**

**Table 2: Robustness Efficiency**

| Image | PSNR (dB) | SSIM | Security Value | Mutual Information |
|---|---|---|---|---|
| Lena | 36.74 | 0.9827 | 0.003742 | 0.287313 |
| Fishing Boat | 36.68 | 0.9879 | 0.005608 | 0.287313 |
| Peppers | 39.11 | 0.9850 | 0.002643 | 0.287313 |
| Baboon | 37.68 | 0.9943 | 0.002210 | 0.248601 |
| USAir Force | 34.31 | 0.9798 | 0.011429 | 0.277865 |

| Attacks | Bit Error Rate | |
|---|---|---|
| | Detector I | Detector II |
| Median Filtering | 0.1992 | 0.0117 |
| LPF | 0.3516 | 0.0625 |
| Histogram | 0.1758 | 0.2539 |
| Cropping | 0.0273 | 0.1094 |
| Inverting | 0 | 0.0078 |
| Edge Encoder | 0.0078 | 0.0117 |
| Range [up-215 low-25] | 0.0039 | 0.0117 |
| Gaussian Filtering | 0.3398 | 0.0313 |
| Add Noise [pix-10% amnt-20%] | 0 | 0.0352 |
| Scaling [256-128-256] | 0.3047 | 0.1563 |
| Erode | 0.2695 | 0.1133 |
| Dilate | 0.2383 | 0.1133 |
| Gamma Correction | 0 | 0 |
| Edge | 0.2773 | 0.2695 |

**Table 3: Comparison of Performance against Different Image Impairments.**

| Attacks | Normalized Cross-Correlation (NCC) | | |
|---|---|---|---|
| | Cox's Method | Nafornita's Method | Proposed Method |
| Median, size 3x3 | 0 | 0.96 | 1 |
| Intensity Adjustment | 0 | 1 | 1 |
| Scaling 256>128>256 | 0 | 0.53 | 0.84 |
| Crop, ½ | 0 | 0.45 | 0.97 |
| AWGN, SNR=11.4dB | 0.39 | 0.89 | 0.69 |
| JPEG, Q = 20 (CR = 15) | 0.09 | 0.89 | 0.92 |
| JPEG, Q = 25 (CR = 12.8) | 0.18 | 0.89 | 0.93 |
| JPEG, Q = 50 (CR = 8.3) | 0.67 | 1 | 1 |
| JPEG, Q = 75 (CR = 5.5) | 0.99 | 1 | 1 |

## 4. Conclusion

The digital watermarking can be thought as digital communication scheme where an auxiliary message is embedded in digital multimedia signals and are available where ever the later signals move. Therefore, the detection reliability is significantly enhanced by embedding rather transmitting the same watermark through different sub channels (bands). Thus, this diversity technique can give very good results in detecting the watermark, considering the fact that many watermark attacks are more appropriately modeled as fading like [11].

In this paper, we have critically analyzed pixel wise masking of edges and textures as a significant approach that has considerable impact on imperceptibility, detection reliability and data embedding capacity in HVS watermarking.

It is found that data embedding in LH, HL and HH sub bands along with optimized detector I (better against high frequency attacks) and II (better against low frequency attacks) structures offers better resiliency against various types of image distortions. The proposed HVS watermarking scheme also offers visual and statistical invisibility and better security of the hidden data.

The performance of the proposed method is compared and hence found much better with the two state-of-the-art frequency domain watermarking techniques i.e. Cox [2] and Nafornita's works [8].

Future work should concentrate into better use of the HVS properties as well as coding the watermark bits. Although the reported results are based on images, the same conclusions can be extended for other kind of data like audio, music, video etc.

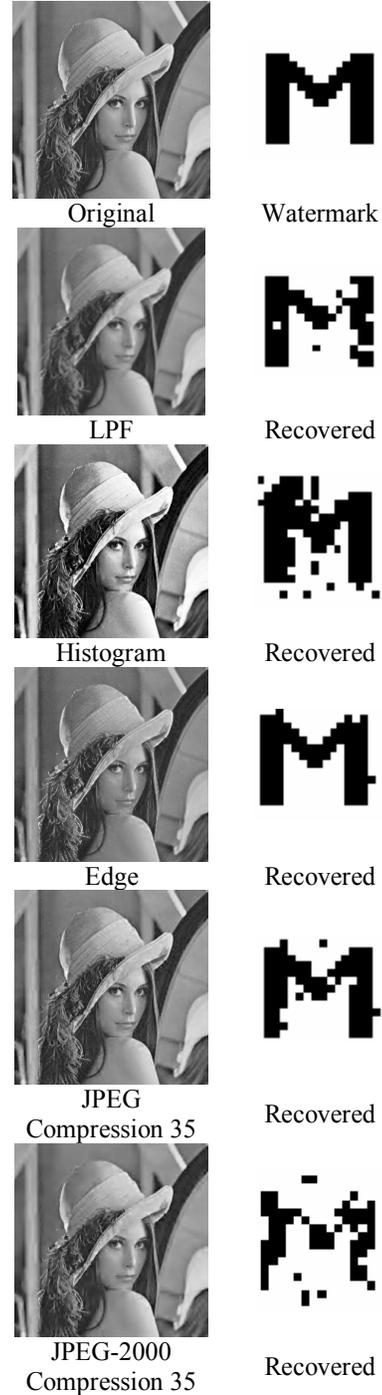

Original — Watermark
LPF — Recovered
Histogram — Recovered
Edge — Recovered
JPEG Compression 35 — Recovered
JPEG-2000 Compression 35 — Recovered

**Fig. 3: Robustness Efficiency under Various Impairements**